\documentclass[journal]{IEEEtran}
\usepackage{cite}
\usepackage[left=1.608cm,right=1.608cm,top=1.8cm]{geometry}

\makeatletter
\renewcommand{\fnum@figure}{Fig. \thefigure} 
\newcommand\identity{1\kern-0.25em\text{l}} 
\makeatother
\usepackage{balance}
\usepackage{amsfonts}
\usepackage[pdftex]{graphicx}
\graphicspath{{../pdf/}{../jpeg/}}
\DeclareGraphicsExtensions{.pdf,.jpeg,.png}
\usepackage[cmex10]{amsmath}
\usepackage{mathabx}
\usepackage{array}
\usepackage{mdwmath}
\usepackage{mdwtab}
\usepackage{eqparbox}
\usepackage{url}
\usepackage{color,soul}
\usepackage{multirow} 
\usepackage{dirtytalk} 
\usepackage[utf8]{inputenc}
\usepackage[english]{babel}
\usepackage{lettrine}
\usepackage{caption}
\usepackage{subcaption}
\usepackage{adjustbox}
\usepackage{amssymb}
\usepackage{xcolor}
\usepackage{algorithm}
\usepackage{algpseudocode}
\newcommand\norm[1]{\lVert#1\rVert} 

\algrenewcommand\algorithmicrequire{\textbf{Input:}} 
\algrenewcommand\algorithmicensure{\textbf{Output:}} 
\usepackage[nocomma]{optidef} 

\usepackage[acronym]{glossaries}
\newacronym{OFDM}{OFDM}{orthogonal frequency division multiplexing}
\newacronym{CP}{CP}{cyclic prefix}
\newacronym{JCAS}{JCAS}{joint communication and  radar sensing}
\newacronym{RF}{RF}{radio frequency}
\newacronym{QAM}{QAM}{quadrature amplitude modulation}
\newacronym{MUSIC}{MUSIC}{multiple signal classification}
\newacronym{ESPRIT}{ESPRIT}{estimation of signal parameters via rotational invariance technique}
\newacronym{CS}{CS}{compressive sensing}
\newacronym{FPS}{FPS}{Fourier projection-slice}
\newacronym{LMMSE}{LMMSE}{linear minimum mean square error}
\newacronym{LS}{LS}{least square }
\newacronym{MMSE}{MMSE}{minimum mean square error}
\newacronym{CE}{CE}{channel estimation}
\newacronym{RMSE}{RMSE}{root mean square error}
\newacronym{TDD}{TDD}{Time Division Duplexing}
\newacronym{FDD}{FDD}{Frequency Division Duplexing}
\newacronym{UL}{UL}{Uplink}
\newacronym{DL}{DL}{deep learning}
\newacronym{BS}{BS}{base station}
\newacronym{MRC}{MRC}{Maximum Ratio Combining}
\newacronym{CSI}{CSI}{channel state information}
\newacronym{ACI}{ACI}{adjacent channel interference}
\newacronym[longplural={User Equipments}]{UE}{UE}{User Equipment}
\newacronym{3GPP}{3GPP}{3rd generation partnership project }
\newacronym{LTE}{LTE}{long term evolution}
\newacronym{GT}{GT}{ground-truth}
\newacronym{SFT}{SFT}{sparse Fourier transform}
\newacronym{ED}{ED}{Energy Detection}
\newacronym{ASK}{ASK}{Amplitude-shift keying}
\newacronym{PSK}{PSK}{Phase-shift keying}
\newacronym{ZCP}{ZCP}{Zadoff-Chu precoding}
\newacronym{ISM}{ISM}{industrial scientific and medical}
\newacronym{SNR}{SNR}{signal-to-noise ratio}
\newacronym{IDFT}{IDFT}{inverse discrete Fourier transform }
\newacronym{DFT}{DFT}{discrete Fourier transform }
\newacronym{IFFT}{IFFT}{inverse fast Fourier transform }
\newacronym{FFT}{FFT}{fast Fourier transform }
\newacronym{ISI}{ISI}{inter symbol interference}
\newacronym{ICI}{ICI}{inter carrier interference}
\newacronym{MIMO}{MIMO}{multiple input multiple output}
\newacronym{PAPR}{PAPR}{peak-to-average power ratio}
\newacronym{GLRT}{GLRT}{Generalized Likelihood Ratio Test}
\newacronym{i.i.d}{i.i.d}{independent and identically distributed}
\newacronym{PSNR}{PSNR}{peak signal to noise ratio}
\newacronym{AWGN}{AWGN}{additive white Gaussian noise}
\newacronym{ZCT}{ZCT}{Zaddof-Chu transform}
\newacronym{HPA}{HPA}{high power amplifier}
\newacronym{RFI}{RFI}{Radio Frequency Interferences}
\newacronym{2D}{2D}{two-dimensional}
\newacronym{1D}{1D}{one-dimensional}
\newacronym{MSE}{MSE}{mean square error}
\newacronym{BER}{BER}{bit error rate}
\newacronym{DCP}{DCP}{disciplined convex programming}

\newacronym{ISAC}{ISAC}{integrated sensing and communication}
\newacronym{ULA}{ULA}{uniform linear array}
\newacronym{HPSD}{HPSD}{Hermittian positive semidefinite}
\newacronym{SINR}{SINR}{signal to interference plus noise ratio}
\newacronym{SCA}{SCA}{successive convex approximation}
\newacronym{SDP}{SDP}{semidefinite programming}
\newacronym{CDF}{CDF}{cumulative distribution function}
\newacronym{NOMA}{NOMA}{non-orthogonal multiple access}

\begin{document}

\bstctlcite{IEEEexample:BSTcontrol}

\title{\Huge Interference Reduction Design for Improved Multitarget Detection in ISAC Systems}
\author{
\uppercase{Mamady Delamou}\authorrefmark{1},
\uppercase{Guevara Noubir\authorrefmark{2}, 
\uppercase{Shuping Dang\authorrefmark{3} and
El Mehdi Amhoud}\authorrefmark{1}}
}

\author{\IEEEauthorblockN{Mamady  Delamou\IEEEauthorrefmark{1} and El Mehdi Amhoud\IEEEauthorrefmark{1}\\
}
\IEEEauthorblockA{\IEEEauthorrefmark{1}College of Computing, Mohammed VI Polytechnic University, Ben Guerir, 43150, Morocco,\\ 
} 
{\{mamady.delamou, elmehdi.amhoud\}@um6p.ma}}

\maketitle
\begin{abstract}
The advancement of wireless communication systems toward 5G and beyond is spurred by the demand for high data rates, exceedingly dependable low-latency communication, and extensive connectivity that aligns with sensing requisites such as advanced high-resolution sensing and target detection. Consequently, embedding sensing into communication has gained considerable attention. In this work, we propose an alternative approach for optimizing integrated sensing and communication (ISAC) waveform for target detection by concurrently maximizing the power of the communication signal at an intended user and minimizing the multi-user and sensing interference. We formulate the problem as a non-disciplined convex programming (NDCP) optimization and we use a distribution-based approach for interference cancellation. Precisely, we establish the distribution of the communication signal and the multi-user communication interference received by the intended user, and thereafter, we establish that the sensing interference can be distributed as a centralized Chi-squared if the sensing covariance matrix is idempotent. We design such a matrix based on the symmetrical idempotent property. Additionally, we propose a disciplined convex programming (DCP) form of the problem, and using successive convex approximation (SCA), we show that the solutions can reach a stable waveform for efficient target detection. Furthermore, we compare the proposed waveform with state of the art radar-communication waveform designs and demonstrate its superior performance by computer simulations.
\end{abstract}

\section{Introduction}
\indent While the implementation of 5G networks is still underway, certain fundamental principles have been identified as the cornerstone of the network, massive machine-type communications (mMTC) for the deployment of a million low-power and short-range devices, ultra-reliable low latency communications (URLLC) for one-millisecond latency and enhanced mobile broadband (eMBB) encompasses mobile applications like augmented reality and video streaming, necessitating swift connections and high data rates \cite{9500443}.\\
\indent The upcoming evolution of wireless networks, commonly referred to as 6G, anticipates a transformative paradigm that emphasizes dependable and swift global connectivity. This envisioned paradigm seeks to establish a network infrastructure capable of delivering not only fast and reliable connections worldwide but also ensuring ultra-low latency. Simultaneously, this next-generation network design is aligned with the growing demands of sensing applications. The integration of sensing requirements within this framework facilitates real-time data transmission and accurate object detection. It goes further to enhance hardware effectiveness by employing a unified waveform, serving the dual functions of \gls{ISAC} \cite{bazzi2023integrated}. Among the various \gls{ISAC} approaches presented in preceding years, the prominent concepts are the coexistence and the dual-functional model. In the former, both communication and sensing entail the transmission of distinct signals that intersect either in the time domain, frequency domain, or both. In contrast, the latter approach merges both signals by employing a unified waveform for both functionalities, effectively integrating communication and sensing within the same signal structure \cite{9200993}. \\
\indent Numerous methodologies have been proposed for single-antenna radar systems \cite{6784117,8528529,8048004,10328976,10200157}. However, leveraging multi-antenna processing has shown substantial potential for enhancing radar performance by delving into \gls{MIMO} radar waveform design \cite{4350230,8579200}.\\
\indent Various studies have already explored waveform optimization, considering diverse aspects. For instance, the work documented in \cite{bazzi2023integrated} addresses the challenge of designing waveforms for dual-functional radar and communication (DFRC) systems, focusing on adjustable peak-to-average power ratio (PAPR). This approach aims to minimize interference in multi-user communication scenarios while adhering to a similarity constraint essential for radar chirp signal characteristics. In addition, the study in \cite{9200993} introduces separated and shared antenna deployments aiming to optimize the weighted sum rate and probing power at the target location. Another approach is presented in \cite{9839026}, where the scheme adapts \gls{NOMA}-inspired interference cancellation, treating part of the sensing signal as virtual communication signals canceled at each user. Moreover, in \cite{8386661}, the authors derived optimal transmission waveforms aimed at minimizing multi-user interference across various radar sensing criteria. On the other hand, in \cite{8288677}, the authors focused on addressing the challenge of aligning a desired sensing beampattern while considering the necessities of the communication \gls{SINR}. Furthermore, in \cite{9124713}, the authors presented a specialized sensing signal and conducted combined optimization with communication signals to achieve maximum degrees of freedom (DoFs) for radar sensing. Meanwhile, \cite{10086626} further investigated the efficacy of the dedicated sensing signal, demonstrating that eliminating sensing interference at the receiver substantially improves the \gls{ISAC} system.\\
\indent However, the constraint of fast and accurate detection poses a primary challenge for most of the previous approaches. In this work, we introduce a new method to combat the sensing complexity and sensing accuracy challenges noticed in several works, namely, the ones proposed in \cite{10086626,9839026,9124713}. Our formulation relies on the \gls{CDF} of the communication signal, the inter-user communication interference, and the sensing interference received by a specific intended user. Upon comparison with similar state of the art methods, our proposed scheme demonstrates the capability to attain an optimal waveform while significantly reducing complexity.\\
In sum, our main contributions in this paper are summarized as follows:
\begin{itemize}
    \item We propose a distribution-based interference reduction for target detection in \gls{ISAC}. We show that the multi-user communication interference and the sensing interference can be reduced based on their respective distributions.
    \item Furthermore, we introduce a \gls{DCP} formulation of the non-convex problem and demonstrate that the solutions lead to an optimal waveform suitable for efficient target sensing.
    \item Finally, we compare our scheme to similar state of the art methods, illustrating the superior performance of our solution.
\end{itemize}

The remainder of the paper is organized as follows: In Section II, we introduce the system model and formulate the non-convex optimization problem. Thereafter, in Section III,  we provide a comprehensive breakdown of the transformational steps to solve the problem. Section IV presents the simulation results, and finally, in Section V, we conclude and set forth our perspectives.\\
\indent In the sequel, we use the following notations. The superscripts,$^H$ and $^T$ denote the conjugate transposition and the transposition,
respectively. Also, $\lvert \cdot \rvert$ takes the modulus of a complex number and $\lVert \cdot \rVert$, Tr($\cdot$), and Rk($\cdot$) return the Frobenius norm, the trace, the rank of a matrix. In addition, Prob($\cdot$) denotes the probability, and $\mathbf{I}_n$ is the identity matrix of order $n$. We also recall the upper incomplete gamma function and the lower incomplete gamma function defined as 
\mbox{$\Gamma(q, p)=\int_p^{\infty} t^{q-1} e^{-t} d t$}, and  \mbox{$\gamma(q, p)=\int_0^p t^{q-1} e^{-t} d t$}, respectively.

\section{System Model and Problem Formulation}
\subsection{System Model}
     \begin{figure}[t!] 
    \centering
    \includegraphics[width=3.5in]{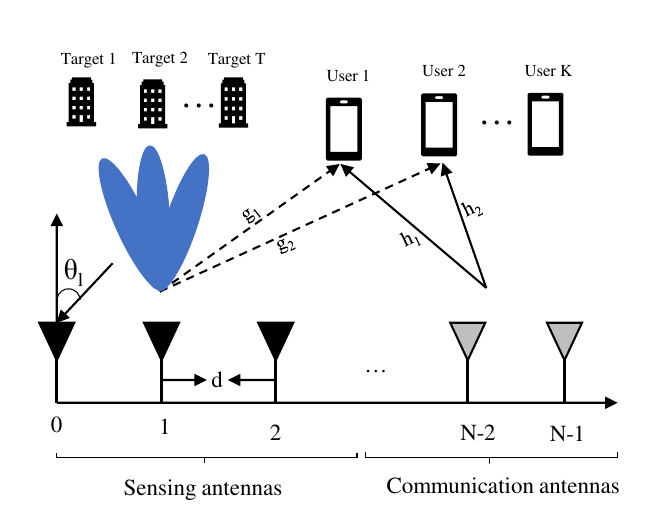}
    \caption{System model}
    \label{system_mod_1}
\end{figure}

As shown in Fig. 1, we consider a multi-antenna \gls{ISAC} system, which consists of a dual-functional $N$-antenna \gls{BS} split into $N_c$ communication antennas and $N_s$ sensing antennas such as $N = N_c + N_s$, $K$ single-antenna communication users indexed by $\mathcal{K} = \{1,\dots, K\}$, and sensing targets denoted by $\mathcal{T} = \{1,\dots,T\}$. The \gls{BS} transmits the communication signals $\mathbf{w}_i s_i$, $\forall i \in \mathcal{K}$ and a dedicated sensing signal $\mathbf{r} \in \mathbb{C}^{N_s\times1}$, where $\mathbf{w}_i \in \mathbb{C}^{N_c\times1}$ denotes the beamformer for transmitting the information symbol $s_i \in \mathbb{C}$ to user $i$. In addition, the total power budget for either deployment $P_t$ is shared between the sensing power $P_s$ and communication power $P_c$ \cite{9200993}. The information symbols $s_i, i \in \mathcal{K}$ are assumed to be independent as well as with zero mean and unit power. Both the communication precoders and the radar signals are designed to fulfill the dual function. The received signal at user $k$ can be expressed as \cite{9839026}:
\begin{equation}
y_k=\underbrace{\mathbf{h}_k^H \mathbf{w}_k s_k}_{\text {useful signal}} + \underbrace{\mathbf{h}_k^H \sum_{i \in \mathcal{K}, i \neq k} \mathbf{w}_i s_i}_{\text {inter-user interference }} + \underbrace{\mathbf{g}_k^H\mathbf{r}}_{\text {sensing interference }}  + \underbrace{n_k}_{\text {noise}},
\end{equation}
where $\mathbf{h}_k \in \mathbb{C}^{N_c\times1}$, $\mathbf{g}_k \in \mathbb{C}^{N_s\times1}$, $\forall k \in \mathcal{K}$ denotes the \gls{BS} communication and sensing channel to user-$k$, respectively. $n_k \in \mathbb{C}$, $\forall k \in \mathcal{K}$ denotes the circularly symmetric complex Gaussian noise with variance $\sigma^2_n$.\\
Thus, by denoting $\mathbf{W} = [\mathbf{w}_1,\dots, \mathbf{w}_K]$ the precoder matrix, the achievable rate at user $k$ is given by
\begin{equation}\label{rate}
R_k(W)=\log _2\left(1 + \frac{\left|\mathbf{h}_k^H \mathbf{w}_k\right|^2}{\sum_{i \in \mathcal{K}, i \neq k}\left|\mathbf{h}_i^H \mathbf{w}_i\right|^2 +  \mathbf{g}_k^H \mathbf{C}\mathbf{g}_k +\sigma_n^2}\right),
\end{equation}
where $\mathbf{C} = \mathbb{E}[\mathbf{rr}^H]$ is the covariance matrix of the transmit radar signal and $\mathbf{C}$ is \gls{HPSD}.\\
\indent The transmit beampattern is essentially a representation of how the transmitted signal power is spatially distributed across different angles or directions. The transmit beampattern in a direction of azimuth angle $\theta_l$ is given by
\begin{equation}
   \Tilde{\Phi}(\theta_l) = \mathbf{a}^H(\theta_l)\mathbf{C}\mathbf{a}(\theta_l),
\end{equation}
where $\mathbf{a} = [1, e^{j\frac{2\pi}{\lambda}dsin(\theta_l)}, \dots, e^{j\frac{2\pi}{\lambda}d(N_s-1)sin(\theta_l)}]^T$ denotes the steering vector of the \gls{ULA} at the \gls{BS}.
\subsection{Problem Formulation}
In practice, the desired sensing beampattern is designed according to the sensing requirements \cite{4276989}. In this work, we hypothesize that the radar is working in tracking mode. In this scenario, the sensing system possesses prior knowledge about the targets, the beampattern is anticipated to exhibit prominent peaks in the directions of the identified targets. However, in cases where the sensing system operates without prior target information, an isotropic beampattern is preferred. This means that the power is evenly distributed across all directions.\\
\indent  The optimization aims to minimize the matching error, calculated in the least squares sense, between the transmitted beampattern and the desired sensing beampattern for efficient target detection. This optimization is conducted while adhering to the constraints that the inter-user communication interference and the sensing interference do not exceed a certain threshold, and the communication signal power received at the desired user $k$ is above a certain threshold.\\
\indent From (\ref{rate}), these constraints are mathematically expressed as \mbox{Prob$(\left|\mathbf{h}_i^H \mathbf{w}_i\right|^2 \le \beta_i) \ge \epsilon_i,  \forall i \ne k$}, \mbox{Prob$(\mathbf{g}_k^H \mathbf{C}\mathbf{g}_k  \le \rho) \ge \alpha$} and \mbox{Prob$(\left|\mathbf{h}_k^H \mathbf{w}_k\right|^2 \ge \xi) \ge \nu$}. The resultant optimization problem can be formulated as
\begin{mini!}|s|[2]
    {\mathbf{W},\delta,\mathbf{C}}
    { \sum_{t=1}^T \lvert \delta\Phi(\theta_t)-\Tilde{\Phi}(\theta_t)\rvert^2        \label{eq:eq1}}
    {\label{opt1}}
    {}   
    \addConstraint{Prob(\left|\mathbf{h}_k^H \mathbf{w}_k\right|^2 \ge \xi) \ge \nu \label{opt1:con1}}, 
    \addConstraint{Prob(\mathbf{g}_k^H \mathbf{C}\mathbf{g}_k  \le \rho) \ge \alpha \label{opt1:con2}}, 
    \addConstraint{Prob(\left|\mathbf{h}_i^H \mathbf{w}_i\right|^2 \le \beta_i) \ge \epsilon_i, \forall i \ne k \label{opt1:con3}},      
    \addConstraint{\operatorname{Tr}\left(\mathbf{C}\right)} \le {P_s \label{opt1:con4}}, \addConstraint{\operatorname{Tr}\left(\mathbf{W}\mathbf{W}^H\right)} \le {P_c \label{opt1:con5}},
    \addConstraint{\mathbf{C}} \ge {0 \label{opt1:con6}},
\end{mini!}
where $\Phi(\theta_t)$ is the desired beampattern and $\delta$ is a scaling factor that keeps $\Phi(\theta_t)$ and  $\Tilde{\Phi}(\theta_t)$ within the same scale for better interpretation. Constraints (\ref{opt1:con4}) and (\ref{opt1:con5}) ensure that the transmitting powers for sensing and communication do not exceed the respective allocated budgets. Finally, (\ref{opt1:con6}) indicates that $\mathbf{C}$ is positive semi-definite.
In practical scenarios, the \gls{CSI} accessible to the transmitter is often imperfect due to estimation errors or various factors like quantization. Specifically, we characterize these variations as an additive complex Gaussian noise, thereby modeling the channels as \cite{5403537}:
\begin{equation}\label{channel_distribution}
     \mathbf{h_i}  \sim \mathcal{C N}\left(\mathbf{0}, \sigma_{h_i}^2 \mathbf{I_{N_c}}\right)  ~\mathrm{and}~  \mathbf{g_i}  \sim \mathcal{C N}\left(\mathbf{0}, \sigma_{g_i}^2 \mathbf{I_{N_s}}\right),\forall i,
\end{equation} 
where the variances $\sigma^2_{h_i}$ and $\sigma^2_{g_i}$ indicate the \gls{CSI} quality. \\
 \indent It is worth mentioning that the primary goal is to optimize the system in a way that by controlling the interference, the transmit beampattern can locate the target's azimuth angles. Consequently, the inter-user communication interference and sensing interference are managed probabilistically at predefined levels, $\epsilon$ and $\alpha$, to ensure a harmonious trade-off scenario. Provided that $\mathbf{h}_k$ is an additive complex Gaussian variable, $u_k \triangleq \left|\mathbf{h}_k^H \mathbf{w}_k\right|^2$ is recognized as a Chi-squared random variable $\chi^2$ with degrees of freedom $n_{u_k}$ = 2, a variance $\sigma^2_{u_k} = \frac{\left||\mathbf{w}_k\right||^2 \sigma^2_{h_k}}{2}$, and a noncentrality parameter $s^2_{u_k} = \left|\mathbb{E}[\mathbf{h}_k]\mathbf{w}_k\right|^2 = 0$. As such,
\begin{equation}\label{proba_to_Q_function}
    Prob(\left|\mathbf{h}_k^H \mathbf{w}_k\right|^2 \ge \xi) = Q(\frac{s_{u_k}}{\sigma_{u_k}}, \frac{\sqrt{\xi}}{\sigma_{u_k}}) = Q(0, \frac{\sqrt{\xi}}{\sigma_{u_k}}),
\end{equation}
where $Q(.,.)$ denotes the generalized Marcum’s $Q$-function. By applying (\ref{proba_to_Q_function}) to (\ref{opt1:con3}), we have 
\begin{equation}
    \begin{split}
  Prob(\left|\mathbf{h}_i^H \mathbf{w}_i\right|^2 \le \beta_i) &= 1 - Prob(\left|\mathbf{h}_i^H \mathbf{w}_i\right|^2 \ge \beta_i)\\ 
  & = 1-Q(0, \frac{\sqrt{\beta_i}}{\sigma_{u_i}}).       
    \end{split}
\end{equation}
\indent Provided that $\mathbf{C}$ is \gls{HPSD} and $\mathbf{g_k}$ is a complex vector such that $\mathbf{g_k}\sim \mathcal{C N}\left(\mathbf{0}, \sigma_{g_i}^2 \mathbf{I_{N_s}}\right)$, the \gls{CDF} of $\mathbf{g}_k^H\mathbf{C}\mathbf{g}_k$ can be a Chi-squared of freedom $2m$ ($\chi_{(2m)}^2$) if and only if $\mathbf{C}$ is idempotent of rank $m$ \cite{idempotentAndChisquare}. Let us suppose that $\mathbf{C}$ is idempotent. As a result, $\mathbf{g}_k^H\mathbf{C}\mathbf{g}_k \sim \chi_{(2m)}^2$, therefore, the probability in (\ref{opt1:con2}) can be expressed as $Prob(\mathbf{g}_k^H\mathbf{C}\mathbf{g}_k \le\rho) = F(2m;\rho) = P(m,\frac{\rho}{2})$, with $P(m,\frac{\rho}{2}) =\frac{\gamma(m,\frac{\rho}{2})}{\Gamma(m)}$. $F$ and $P$ denote the cumulative distribution function of $\chi^2$ and  the regularized gamma function, respectively.
Hence, the optimization problem in (\ref{opt1}) becomes 

\begin{mini!}|s|[2]
    {\mathbf{W},\delta,\mathbf{C}}
    { \sum_{t=1}^T \lvert \delta\Phi(\theta_t)-\Tilde{\Phi}(\theta_t)\rvert^2 \label{eq:eq2}}
    {\label{opt2}}
    {}  
    \addConstraint{Q(0, a_k) \ge \nu \label{opt2:con1}}, 
    \addConstraint{P(q,\frac{\rho}{2}) \ge \alpha} 
    \label{opt2:con2},
     \addConstraint{Q(0, b_i) \le 1-\epsilon_i, \forall i \ne k \label{opt2:con3}},
    \addConstraint{\operatorname{Tr}\left(\mathbf{C}\right)} \le {P_s \label{opt2:con4}},        \addConstraint{\operatorname{Tr}\left(\mathbf{W}\mathbf{W}^H\right)} \le {P_c \label{opt2:con5}}, 
    \addConstraint{\mathbf{C^2}} =  {\mathbf{C} \label{opt2:con6}},
    \addConstraint{\operatorname{Rk}(\mathbf{C})} =  {m\label{opt2:con7}},
    \addConstraint{\mathbf{C}} \ge {0 \label{opt2:con8}},
\end{mini!}

where $a_{k}=\frac{\sqrt{\xi}}{\sigma_{u_k}} = \frac{\sqrt{\xi}}{\sqrt{\frac{\norm{\mathbf{w}_k}^2 \sigma^2_{h_k}}{2}}}$, \mbox{$b_i = \frac{\sqrt{\beta_i}}{\sigma_{u_k}}= \frac{\sqrt{\beta_i}}{\sqrt{\frac{\norm{\mathbf{w}_i}^2 \sigma^2_{h_i}}{2}}}$}. The new constraints (\ref{opt2:con6}) and (\ref{opt2:con7}) are added to ensure the previous supposition, which is $\mathbf{C}$ being an idempotent matrix of \mbox{rank $m$}.
\section{Resolution of the Beampattern Optimization}
\subsection{Sensing Covariance Matrix Optimization}
The problem defined in (\ref{opt2}) does not follow the \gls{DCP} rules. Therefore, it can not be solved in its current form using a convex optimization tool like CVX. At one hand, the constraints (\ref{opt2:con1}) and (\ref{opt2:con3}) are function of $\frac{1}{\norm{\mathbf{w}_k}^2}$ which do not respect \gls{DCP} rules. On the other hand, the constraints (\ref{opt2:con6}) and (\ref{opt2:con7}) are not convex. To transform the problem into \gls{DCP}, we have performed a change of variable and \gls{SCA}. These methodologies involve reshaping and expanding the problem to adhere to the guidelines and constraints outlined by \gls{DCP} frameworks.\\
\indent By grouping constraints (\ref{opt2:con6}) and (\ref{opt2:con7}), we obtain the following system
\begin{equation}\label{idempotent}
  \begin{cases}
    \mathbf{C^2} =  \mathbf{C}, \\
    \text{Rk}(\mathbf{C}) =  m, \\
\end{cases}  
\end{equation}
which solutions are idempotent matrices of rank $m$. We solved (\ref{idempotent}) based on a symmetrical idempotent property \cite{idempotent_matrix}. As a matter of fact, for a rectangular matrix $\mathbf{A}$ of dimension $n$ by $m$ and with a rank $m$ ($n \ge m \ge 1$), the matrix $\mathbf{M}$ such that
\begin{equation}\label{idempotent_construction}
    \mathbf{M} = \mathbf{A(A^TA)^{-1}A^T}
\end{equation}
verifies $\mathbf{M^2 = M}$. This suggests that the matrix $\mathbf{A}$ may not be a zero matrix, and it must precisely contain a set of $m$ linearly independent columns. As such, $\mathbf{M}$ is square, of order $n$, and symmetrical. It cannot be the identity matrix of order $n$ since $\mathbf{A}$ is singular, and it cannot be the null matrix. In addition, Rk($\mathbf{M}) = m$ \cite{idempotent_matrix}. Hence, $\mathbf{C}$ can be efficiently constructed using (\ref{idempotent_construction}). 
The fulfillment of constraint (\ref{opt2:con6}) and (\ref{opt2:con7}) become readily achievable through the construction outlined in (\ref{idempotent_construction}) by appropriately selecting the right value of $m$. Moreover, the non-convexity of (\ref{opt2:con1}) and (\ref{opt2:con3}) only relies on $\frac{1}{\norm{\mathbf{w}_i}}, \forall i$. Thoroughly, $\norm{\mathbf{w}_i}$ is convex but $\frac{1}{\norm{\mathbf{w}_i}}$ is not \gls{DCP}. To convexify it, the sub-optimization problem of looking for $\mathbf{W}$ such that (\ref{opt2:con1}) and (\ref{opt2:con3}) are satisfied can be reformulated by using change of variable $v_i = \frac{1}{\norm{\mathbf{w}_i}}$. The rationale behind this change of variable is to shift the non-convex challenge to the level of constraint (\ref{opt2:con5}), which can be expressed as $\sum_{i=1}^\mathcal{K}\frac{1}{v^2_i} \le P_c$ and can be solved using \gls{SCA} as outlined in Algorithm \ref{pseudo_algo1}. This approach allows for a seamless transition, utilizing the transformed problem to establish matrices that comply with the desired conditions, thereby paving the way for enhanced problem-solving strategies. In the following, we consider the first order Marcum’s $Q$-function, $Q_1$, and we have \mbox{$Q_1(0, z) = \frac{\Gamma(1, z^2/2)}{\Gamma(1)}$} and  \mbox{$\Gamma(1,z^2/2) = e^{-z^2/2}$}. Since \mbox{$e^{-z^2/2}$} is log-concave, (\ref{opt2:con3}) does not respect \gls{DCP}. However, it can be solved by the means of \gls{SCA}. We consider the Taylor second-order expansions $\Tilde{f}$ and $\Tilde{g}$ as the surrogate functions for \mbox{$f(z) = e^{-z^2/2}$} and \mbox{$g(v) = \sum_{i=1}^\mathcal{K}\frac{1}{v_i^2}, \forall i \ne k$} at $z_0$ and $v_0$, respectively, with \mbox{$v = (v_i)_{i \ne k}$}. The problem (\ref{opt2}) can be rewritten as
\begin{mini!}|s|[2]
    {\mathbf{v},\delta,\mathbf{C}}
    { \sum_{t=1}^T \lvert \delta\Phi(\theta_t)-\Tilde{\Phi}(\theta_t)\rvert^2 \label{eq:eq3}}
    {\label{opt3}}
    {}    
    \addConstraint{\operatorname{Construct\mathbf{~} \mathbf{C}\mathbf{~} using \mathbf{~} (\ref{idempotent_construction})} \label{opt3:con1}},
    \addConstraint{e^{-a_k^2/2} \ge \nu \label{opt3:con2}}, 
    \addConstraint{P(m,\frac{\rho}{2}) \ge \alpha} \label{opt3:con3},
    \addConstraint{\Tilde{f} \le 1-\epsilon_i, \forall i \ne k \label{opt3:con4}},    
    \addConstraint{\Tilde{g} \le {P_c,  \forall i \in \mathcal{K} \label{opt3:con5}}},
    \addConstraint{\mathbf{C}} \ge {0 \label{opt3:con6}},
\end{mini!}
where $a_{k} = \frac{2v_k\sqrt{\xi}}{\sigma_{h_k}}$, \mbox{$b_i = \frac{2v_i\sqrt{\beta_i}}{\sigma_{h_i}}$}, $\Tilde{f}(z) = \exp \left(-\frac{z_0^2}{2}\right)-z_0\left(z-z_0\right) \exp \left(-\frac{z_0^2}{2}\right)+\frac{\left(z_0^2-1\right)}{2}\left(z-z_0\right)^2 \exp \left(-\frac{z_0^2}{2}\right)$ and $\Tilde{g}(v) = \sum_{i=1}^K \frac{1}{v_i^2}-2\left(\sum_{i=1}^K \frac{1}{v_i^3}\right)\left(v-v_0\right)+3\left(\sum_{i=1}^K \frac{1}{v_i^4}\right)\left(v-v_0\right)^2$.\\
\indent The optimization problem (\ref{opt3}) can be solved through three sequential steps: (1) initially constructing $\mathbf{C}$ by (\ref{idempotent_construction}), (2) the subsequent resolution of the problem (\ref{opt3}) with the resultant $\mathbf{C}$, and (3) verify the convergence condition in (\ref{opt3:con4}) and (\ref{opt3:con5}) using \gls{SCA}. The comprehensive pseudo-algorithm is succinctly summarized in Algorithm \ref{pseudo_algo2}, where $b^k$, $z^k$, $\Omega_k$ and $\omega_k$ denote the values of $b$, $z$ and the weighting factors for $b^k$ and $z^k$ at the $k$th iteration, respectively.
\begin{algorithm}
\caption{SCA Algorithm}\label{pseudo_algo1}
\begin{algorithmic}[1]
\Require{$\Tilde{f}$}
\State Set $k = 0$, initial feasible point $x_0 \in S_0, {\omega_k} \in (0, 1]$.
\State Repeat
\State\hspace{\algorithmicindent} $\hat{x}(x_k) = \arg \min_{x \in S_0} \Tilde{f}(x|x_k)$
\State\hspace{\algorithmicindent}$x_{k+1} = x_{k} + \omega_k*(\hat{x}(x_k)-x_{k})$
\State\hspace{\algorithmicindent}k = k+1
\State Until convergence
\Ensure{$x_k$}
\end{algorithmic}
\end{algorithm}
\begin{algorithm}
    \caption{Symmetric Idempotent Construction + SCA}\label{pseudo_algo2}
    \begin{algorithmic}[1]
    \Require{$P_c$, A}
    \State Construct $\mathbf{C}$ as $\mathbf{M}$ in (\ref{idempotent_construction})
    \State Repeat
    \State\hspace{\algorithmicindent}Solve (\ref{opt3}) using the $\mathbf{C}$ from (\ref{idempotent_construction})
    \State\hspace{\algorithmicindent}$\hat{z}(z^k) = \arg \min_{z \in S_0} \Tilde{f}(z|z^k)$
    \State\hspace{\algorithmicindent}$\hat{b}(b^k) = \arg \min_{b \in S_1} \Tilde{g}(b|b^k)$
    \State\hspace{\algorithmicindent}$z^{k+1} = z^{k} + \omega_k*(\hat{z}(z^k)-z^{k})$
    \State\hspace{\algorithmicindent}$b^{k+1} = b^{k} + \Omega_k*(\hat{b}(b^k)-b^{k})$
    \State\hspace{\algorithmicindent}$k = k+1$
    \State Until $\lVert z^k -  z^{k-1} \rVert \le \zeta_1$ and $\lVert b^k -  b^{k-1} \rVert \le \zeta_2$
    \Ensure{$b^k$, $z^k$}
    \end{algorithmic}
\end{algorithm}
\subsection{Complexity Analysis}
The burden of the computation of (\ref{opt3}) mainly relies on the number of iterations $I_0$ for performing \gls{SCA}, matrix multiplication, and matrix decomposition. The evaluation of the optimization variables in (\ref{opt3}) is a convex \gls{SDP} problem and has a polynomial worst-case complexity \cite{boyd1038003, BOSE2021107985}. Moreover, since $\mathbf{A}$ is $n \times m$ matrix, the matrix $\mathbf{C}$ constructed based on (\ref{idempotent_construction}) is $n \times n$ matrix. The computational complexity of finding $A^TA$ is $\mathcal{O}(m^2\times n)$. Computing $(A^TA)^{-1}$ typically has a complexity of around $\mathcal{O}(m^3)$. Multiplying $\mathbf{A}$ by $(A^TA)^{-1}$ requires $\mathcal{O}(m^2\times n)$ operations. Finally, multiplying the resulting matrix by $A^T$ has a complexity of $\mathcal{O}(n^2\times m)$. Thus, the overall complexity of evaluating the expression $\mathbf{C}$ is dominated by the matrix multiplication involving also the inversion operation and would typically be $\mathcal{O}(2\times  m^2\times n + m\times n^2 + m^3)$. In addition, performing (\ref{opt3:con3}) necessitates a Taylor expansion of order $Rk(C)$, the complexity is $\mathcal{O}(Rk(C)) = \mathcal{O}(m)$. Furthermore, (\ref{opt3:con2}) and (\ref{opt3:con4}) are performed in $\mathcal{O}(K)$, (\ref{opt3:con5}) is an $\mathcal{O}(K)$ operations and (\ref{opt3:con6}) takes $\mathcal{O}(n^3)$ operations. Finally, since $n = N_s$, the complexity of the proposed algorithm is $\mathcal{O}(I_0(2\times K + m + 2\times m^2\times N_s + N_s^2\times m + m^3 + N_s^3))$.
\section{Simulation Results}


\begin{figure*}[ht]
     \centering
     \begin{subfigure}[b]{0.32\textwidth}
         \centering
         \includegraphics[width=\textwidth]{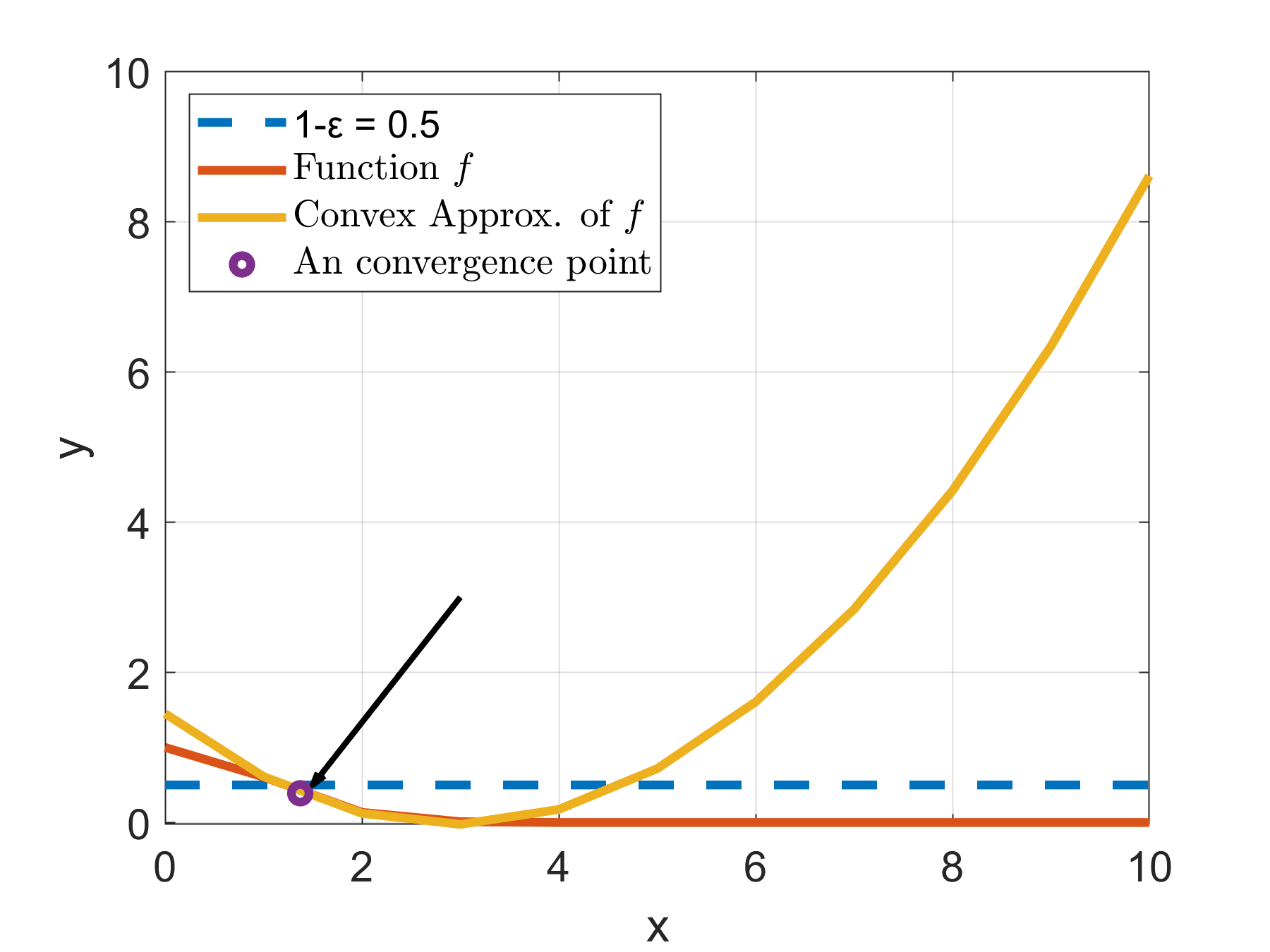}
         \caption{\gls{SCA} of $f(z) = exp(-z^2/2)$ with \mbox{$\epsilon = 0.5$}.}
         \label{sca_expo}
     \end{subfigure}
     \hfill
     \begin{subfigure}[b]{0.32\textwidth}
         \centering
         \includegraphics[width=\textwidth]{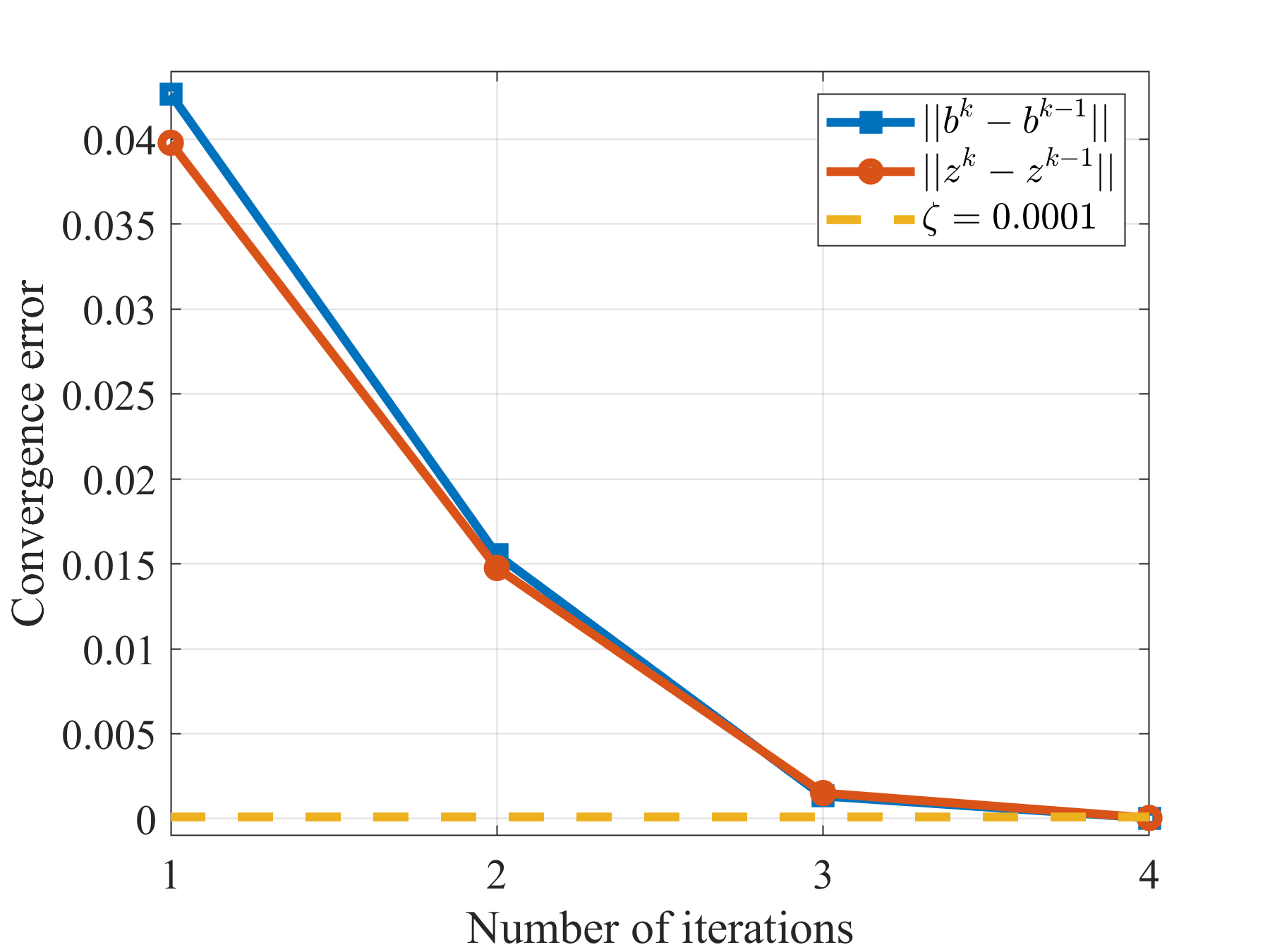}
         \caption{Convergence error of Algorithm \ref{pseudo_algo2}.}
         \label{convergence_limite}
     \end{subfigure}
     \hfill
     \begin{subfigure}[b]{0.32\textwidth}
         \centering
         \includegraphics[width=\textwidth]{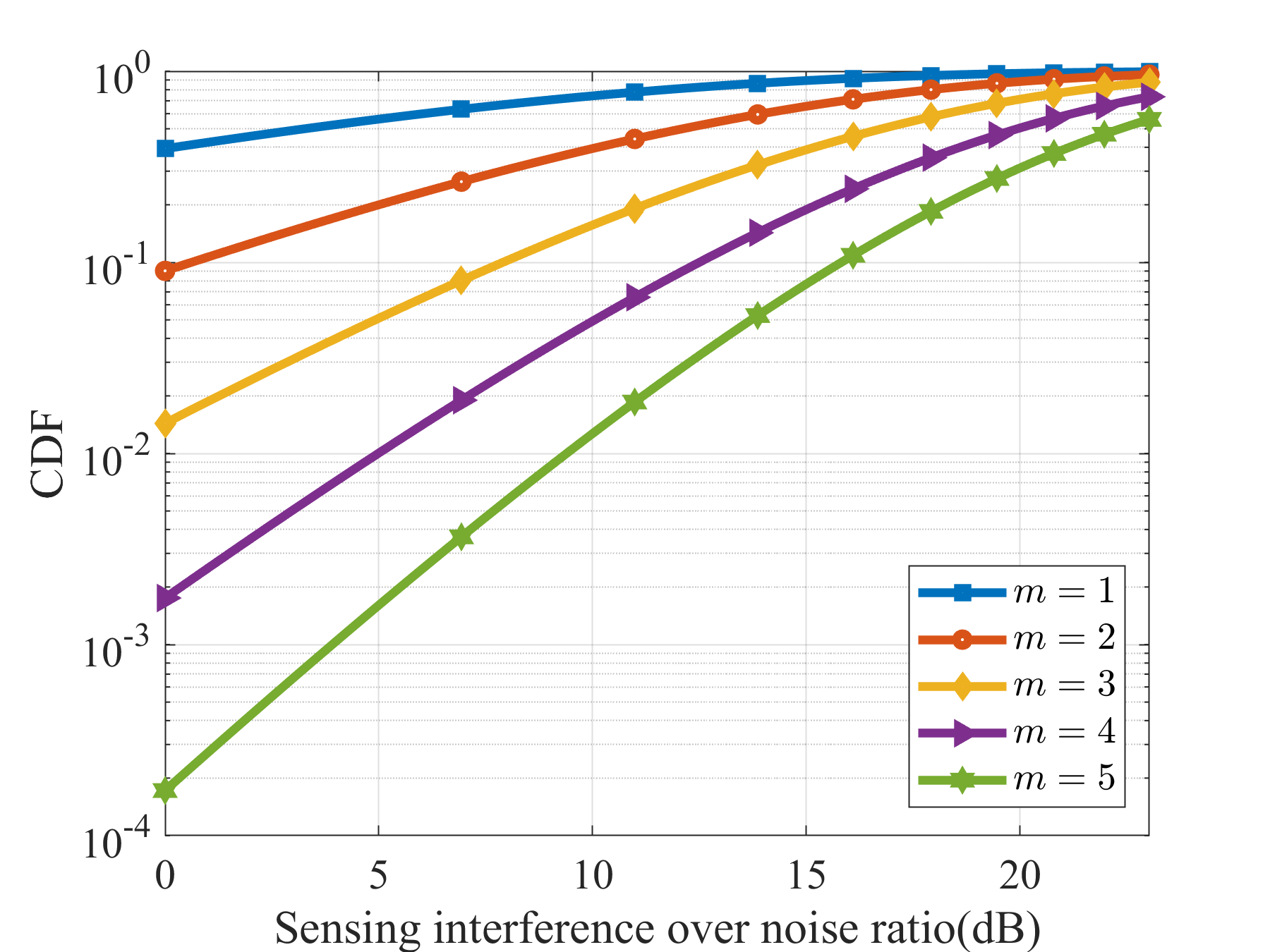}
         \caption{Beampatterns comparison}
         \label{cdf}
     \end{subfigure}
        \caption{Data structure, convergence and sample prediction.}
        \label{}
\end{figure*}
\begin{figure}[t!] 
    \centering
    \includegraphics[width=3.5in]{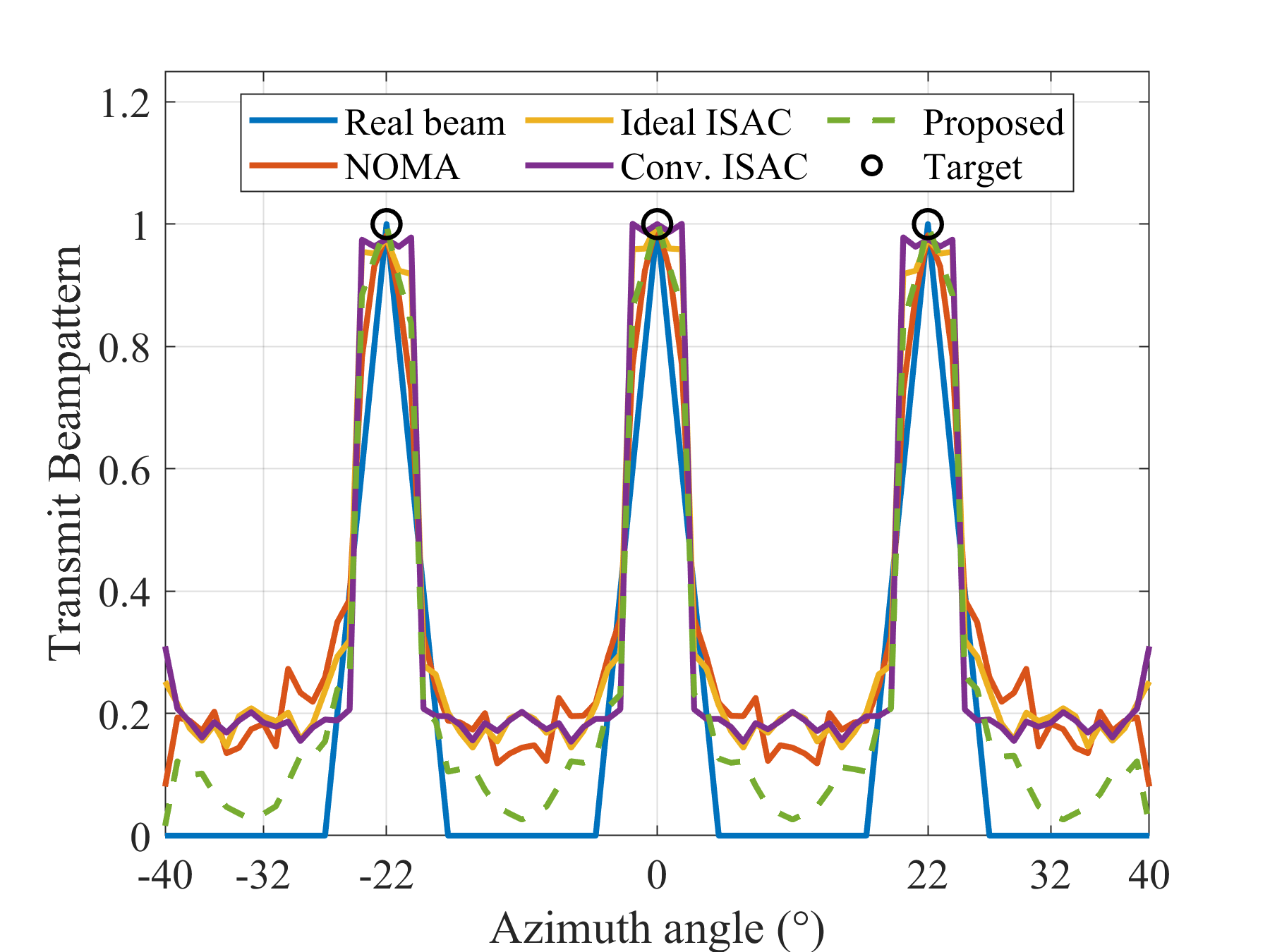}
    \caption{Beampatterns comparison}
    \label{beampattern}
\end{figure}
\begin{table}[ht]
\caption{Execution time (seconds)}
\centering
\begin{tabular}{c c c c}
\hline\hline
NOMA & Ideal ISAC & Conventional ISAC & Proposed \\ [0.5ex] 
\hline
47.19&9.85&9.53&\textbf{5.97}\\
 \\[1ex]
\hline
\end{tabular}
\label{execution_time}
\end{table}
\begin{table}[ht]
\caption{Simulation parameters}
\begin{tabular}{>{\raggedright}p{4cm} c c}
\hline\hline
Parameter & Symbol & Value \\ [0.5ex] 
\hline
Number of antennas & $N$ & 24 \\
Total power & $P_t$ & 10 dBW \\
Number of users & $K$ & 3 \\
Number of targets & $T$ & 3 \\
Target directions & $\phi$ & $[-22^{\circ}, 0^{\circ}, 22^{\circ}]$ \\
Channel error variance & $\sigma_{h_i}^2, \sigma_{g_i}^2$ & 0.05\\
Sensing, communication power & $P_s,P_c$ & $\frac{P_t}{2}$ \\
Sensing, communication antennas & $N_s,N_c$ & $\frac{N}{2}$ \\
Inter-user interference & $frac{\beta_i}{\sigma^2_n}$ & 0.1 \\
Sensing interference& $\frac{\rho}{\sigma^2_n}$&2\\
Communication power threshold &$\frac{\xi}{\beta_i}$& 0.01\\[1ex]
\hline
\end{tabular}
\label{simulation_parameters}
\end{table}
In this section, we present numerical results to characterize the proposed \gls{ISAC} system. Our analysis is based on a dual-functional \gls{BS} featuring a \gls{ULA} antennas with half-wavelength spacing. Given the directions of targets, the desired beampattern is given by 
\begin{equation}
 \Phi(\theta_l)=
 \begin{cases}
    \frac{\theta_l}{\Delta} + 1-\frac{\varphi}{\Delta} ,  \mathrm{~} \theta_l < \varphi,  \mathrm{~}\varphi \in \phi \\
     -\frac{\theta_l}{\Delta} + 1+\frac{\varphi}{\Delta} ,  \mathrm{~} \theta_l > \varphi,  \mathrm{~}\varphi \in \phi \\
    0, \mathrm{~} \theta_l \ge \varphi + \Delta \mathrm{~}\mathrm{or}\mathrm{~}  \theta_l \le \varphi - \Delta\\
\end{cases}. 
\end{equation}
For comparison, we examine the performance of the similar state of the art targets detection schemes, namely, the ideal sensing interference cancellation (Ideal ISAC) \cite{10086626}, the \gls{NOMA}-inspired interference cancellation for \gls{ISAC} \cite{9839026} and the no sensing interference cancellation (Conventional ISAC) \cite{9124713}.
The channels are independent and identically distributed (i.i.d.) as in (\ref{channel_distribution}). Additionally, we assume the following probabilities \(\epsilon_i = 0.5\) for \(i \neq k\), \(\alpha = 0.9\), and \(\nu = 0.9\). The convergence criterion for the \gls{SCA} is \(\zeta = 0.0001\), $\Omega_k = \omega_k = 0.1^k$ and $\Delta = 5^\circ$. See the other simulation parameters in table \ref{simulation_parameters}.\\ 
\indent Within Table \ref{execution_time}, a comprehensive examination of execution times unveils a notable disparity between our proposed solution and the benchmark methods. The efficiency gains observed in our method are largely attributed to the algebraic construction of matrix $\mathbf{C}$. This algebraic approach renders the optimization of $\mathbf{C}$ substantially more effective, contributing to the remarkable speed exhibited by our solution.\\
 \indent In Fig. \ref{sca_expo}, we illustrate as an example, the \gls{SCA} of \mbox{$f(z) \le 1-\epsilon$}, where $f(z) = \exp(-z^2/2)$ and $\epsilon = 0.5$. This graphical representation provides insight into the convergence behavior of the algorithm and a convergence point at which the inequality (\ref{opt3:con4}) is satisfied.\\
\indent In Fig. \ref{convergence_limite}, the graph illustrates the convergence error of the functions $\begin{array}{@{}r@{\;}c@{\;}c@{\;}l@{}} & z & \mapsto & \Tilde{f}(z) \end{array}$ and $\begin{array}{@{}r@{\;}c@{\;}c@{\;}l@{}} & b & \mapsto & \Tilde{g}(b)\end{array}$ as a function of the number of iterations. The errors exhibit a quick reduction with each successive iteration, implying a fast convergence of our algorithm. The rapid convergence throughout iterations attests to the algorithm's effectiveness in quickly approaching an optimal solution.\\
\indent Fig. \ref{cdf} presents the \gls{CDF} of the power of the received sensing interference at an intended user. The graphic shows that The lower the rank of the matrix $\mathbf{C}$, the higher the reduction in sensing interference.\\
\indent In Fig. \ref{beampattern}, we conducted a comprehensive analysis, comparing the transmit beampatterns resulting from various schemes. All the approaches demonstrate the capability to achieve dominant peaks in the desired target directions. However, the proposed scheme surpasses the conventional ISAC and the ideal ISAC and achieves a minimal error compared to NOMA. This performance superiority is particularly evident in the noteworthy gain achieved by the proposed scheme, effectively reducing the secondary lobes. The comparison reveals that the proposed scheme excels in minimizing power leakage in undesired directions.
\section{Conclusion}
In this work, we introduced a novel method that aims at optimizing the \gls{ISAC} beampattern for target detection. Our approach focuses on designing the signal for a specific user to mitigate interference stemming from multi-user communication and sensing for efficient target detection. Additionally, we illustrate that sensing-induced interference can adhere to a centralized Chi-squared distribution while upholding the idempotent nature of the sensing covariance matrix. Our methodology revolves around constructing this matrix by leveraging its symmetrical idempotent properties. Furthermore, we presented a \gls{DCP} formulation for the originally non-convex problem. This formulation demonstrates that the proposed solution yields an efficient target detection and outperforms several existing \gls{ISAC} target detection approaches. In our future work, we want to propose a design for enhanced communication and extend the directional sensing to omnidirectional sensing.
\balance
\bibliographystyle{IEEEtran}
\bibliography{biblio_traps_dynamics}
\vspace{12pt}
\end{document}